\documentclass[preprint,12pt]{article}

\usepackage{amsmath,amssymb,array,calc,rotating,epsfig,psfrag, amscd, datetime}

\usepackage{color}
\usepackage[
      colorlinks=true,
      urlcolor=blue,    
      filecolor=blue,     
      citecolor=red,
      pdfstartview=FitV,
       bookmarksopen=true    
      ]{hyperref}
\usepackage[left=3cm,top=2.5cm,right=3cm,nohead]{geometry}
\numberwithin{equation}{section}

\newcommand{\nc}{\newcommand}

\definecolor{cardinal}{rgb}{0.6,0,0}
\definecolor{darkgreen}{rgb}{0,0.5,0}
\definecolor{golden}{rgb}{0.92, 0.7, 0}
\definecolor{midnight}{rgb}{0, 0, 0.5}
\definecolor{darkblue}{rgb}{0.2, 0, 0.8}

\nc{\ra}{\rightarrow} 
\nc{\lra}{\leftrightarrow} 
\nc{\Ra}{\Rightarrow} 
\nc{\LRa}{\Leftightarrow} 
\nc{\blp}{{\big (}}
\nc{\brp}{{\big )}}
\nc{\Blp}{{\Big (}}
\nc{\Brp}{{\Big )}}
\nc{\bglp}{{\bigg (}}
\nc{\bgrp}{{\bigg )}}
\nc{\Bglp}{{\Bigg (}}
\nc{\Bgrp}{{\Bigg )}}
\nc{\slb}{{\rm [}}
\nc{\srb}{{\rm ]}}
\nc{\bslb}{{\rm \big [}}
\nc{\bsrb}{{\rm \big ]}}
\nc{\Bslb}{{\rm \Big [}}
\nc{\Bsrb}{{\rm \Big ]}}

\def\al{\alpha}

\def\eps{\epsilon}
\nc{\veps}{\varepsilon}
\def\gam{\gamma}

\def\lam{\lambda}
\def\om{\omega}

\nc{\vphi}{\varphi}
\def\tha{\theta}

\def\sig{\sigma}

\def\Gam{\Gamma}

\def\Lam{\Lambda}
\def\Om{\Omega}
\def\Sig{\Sigma}

\def\coeff#1#2{\relax{\textstyle {#1 \over #2}}\displaystyle}
\def\sl{\frak{sl}}

\def\sp{\frak{sp}}

\nc{\myvspace}{\rule[-1em]{0pt}{2.5em}}
\nc{\bea}{\begin{eqnarray}}
\nc{\eea}{\end{eqnarray}}
\nc{\be}{\begin{equation}}
\nc{\ee}{\end{equation}}
\nc{\barr}{\begin{array}}
\nc{\earr}{\end{array}}

\nc{\co}{{\cal o}}

\nc{\cA}{{\cal A}}
\nc{\cB}{ \cal B}

\def\cD{{\cal D}}

\nc{\cF}{{\cal F}}
\nc{\cG}{{\cal G}}

\def\cI{{\cal I}}

\def\cK{{\cal K}}
\nc{\cL}{{\cal L}}
\nc{\cM}{{\cal M}}

\def\cN{{\cal N}}

\def\cO{{\cal O}}

\nc{\cQ}{{\cal Q}}
\nc{\cR}{{\cal R}}
\def\cS{{\cal S}}
\def\cT{{\cal T}}

\def\cV{{\cal V}}
\def\cV{{\cal V}}
\def\cW{{\cal W}}

\def\cZ{{\cal Z}}
\nc{\cQd}{\cQ^{\dagger}}
\nc{\cRd}{\cR^{\dagger}}
\nc{\BB}{{\mathbb B}}
\nc{\CC}{{\mathbb C}}
\nc{\DD}{{\mathbb D}}
\nc{\EE}{{\mathbb E}}
\nc{\FF}{{\mathbb F}}
\nc{\GG}{{\mathbb G}}
\nc{\HH}{{\mathbb H}}
\nc{\JJ}{{\mathbb J}}
\nc{\MM}{{\mathbb M}}
\nc{\RR}{{\mathbb R}}
\nc{\PP}{{\mathbb P}}
\nc{\QQ}{{\mathbb Q}}
\nc{\UU}{{\mathbb U}}
\nc{\ZZ}{{\mathbb Z}}
\nc{\calone}{{\mathbb 1}}

\nc{\half}{\coeff{1}{2}}
\nc{\quarter}{\coeff{1}{4}}
\nc{\del}{\partial}

\nc{\delbar}{\bar\partial}
\nc{\thalf}{\frac{t}{2}}
\nc{\Spin}{\operatorname{Spin}}
\nc{\SO}{\operatorname{SO}}

\nc{\Sp}{{\rm Sp}}
\nc{\com}[2]{{ \left[ #1, #2 \right] }}
\nc{\acom}[2]{{ \left\{ #1, #2 \right\} }}
\nc{\rr}{\rightarrow}
\nc{\p}{\partial}
\nc{\LT}{{\LL_\T}}
\nc{\Tr}{{\rm Tr}}
\nc{\tr}{{\rm tr}}
\nc{\Adag}{A^{\dagger}}
\nc{\AdagI}{A^{\dagger I}}
\nc{\AdagJ}{A^{\dagger J}}
\nc{\AdagK}{A^{\dagger K}}
\nc{\AdagL}{A^{\dagger L}}
\nc{\AdagM}{A^{\dagger M}}
\nc{\Bdag}{B^{\dagger}}
\nc{\BdagI}{B^{\dagger}_I}
\nc{\BdagJ}{B^{\dagger}_J}
\nc{\BdagK}{B^{\dagger}_K}
\nc{\BdagL}{B^{\dagger}_L}
\nc{\BdagM}{B^{\dagger}_M}
\nc{\Cdag}{C^{\dagger}}
\nc{\CdagI}{C^{\dagger I}}
\nc{\CdagJ}{C^{\dagger J}}
\nc{\CdagK}{C^{\dagger K}}
\nc{\Ddag}{D^{\dagger}}
\nc{\DdagI}{D^{\dagger I}}
\nc{\DdagJ}{D^{\dagger J}}
\nc{\DdagK}{D^{\dagger K}}
\nc{\bva}{\breve{a}}
\nc{\bvb}{\breve{b}}
\nc{\bvc}{\breve{c}}
\nc{\bvd}{\breve{d}}
\nc{\bve}{\breve{e}}
\nc{\bvf}{\breve{f}}
\nc{\bvg}{\breve{g}}
\nc{\bvh}{\breve{h}}
\nc{\bvi}{\breve{i}}
\nc{\bvj}{\breve{j}}
\nc{\bvk}{\breve{k}}
\nc{\bvl}{\breve{l}}
\nc{\bvm}{\breve{m}}
\nc{\bvn}{\breve{n}}
\nc{\bvo}{\breve{o}}
\nc{\bvp}{\breve{p}}
\nc{\brvq}{\breve{q}}
\nc{\bvr}{\breve{r}}
\nc{\bvs}{\breve{s}}
\nc{\bvt}{\breve{t}}
\nc{\bvu}{\breve{u}}
\nc{\bvv}{\breve{v}}
\nc{\bvw}{\breve{w}}
\nc{\bvx}{\breve{x}}
\nc{\bvy}{\breve{y}}
\nc{\bvz}{\breve{z}}

\nc{\bvA}{\breve{A}}
\nc{\bvB}{\breve{B}}
\nc{\bvC}{\breve{C}}
\nc{\bvD}{\breve{D}}
\nc{\bvE}{\breve{E}}
\nc{\bvF}{\breve{F}}
\nc{\bvG}{\breve{G}}
\nc{\bvH}{\breve{H}}
\nc{\bvI}{\breve{I}}
\nc{\bvJ}{\breve{J}}
\nc{\bvK}{\breve{K}}
\nc{\bvL}{\breve{L}}
\nc{\bvM}{\breve{M}}
\nc{\bvN}{\breve{N}}
\nc{\bvO}{\breve{O}}
\nc{\bvP}{\breve{P}}
\nc{\bvQ}{\breve{Q}}
\nc{\bvR}{\breve{R}}
\nc{\bvS}{\breve{S}}
\nc{\bvT}{\breve{T}}
\nc{\bvU}{\breve{U}}
\nc{\bvV}{\breve{V}}
\nc{\bvcV}{\breve{\cV}}
\nc{\bvW}{\breve{W}}
\nc{\bvX}{\breve{X}}
\nc{\bvY}{\breve{Y}}
\nc{\bvZ}{\breve{Z}}

\nc{\ttha}{\tilde{\theta}}
\nc{\ttau}{\tilde{\tau}}
\nc{\tTha}{\tilde{\Theta}}
\nc{\tphi}{\tilde{\phi}}
\nc{\tsig}{\tilde{\sig}}
\nc{\tom}{\widetilde{\om}}
\nc{\tOm}{\widetilde{\Om}}
\nc{\tlam}{\widetilde{\lam}}
\nc{\tLam}{\tilde{\Lam}}
\nc{\tSig}{\widetilde{\Sig}}
\nc{\tPhi}{\tilde{\Phi}}
\nc{\tPhibar}{\ol{\tPhi}}
\nc{\tPi}{\widetilde{\Pi}}
\nc{\tpsi}{\widetilde{\psi}}
\nc{\tPsi}{\tilde{\Psi}}
\nc{\tgam}{\widetilde{\gam}}
\nc{\tGam}{\widetilde{\Gam}}
\nc{\tzeta}{\tilde{\zeta}}
\nc{\tZeta}{\tilde{\Zeta}}
\nc{\teta}{\widetilde{\eta}}
\nc{\teps}{\tilde{\eps}}
\nc{\tveps}{\tilde{\veps}}
\nc{\tEta}{\tilde{\Eta}}
\nc{\tchi}{\tilde{\chi}}
\nc{\tChi}{\tilde{\Chi}}
\nc{\txi}{\tilde{\xi}}
\nc{\tXi}{\widetilde{\Xi}}
\nc{\tnu}{\tilde{\nu}}
\nc{\tmu}{\tilde{\mu}}

\nc{\tb}{\tilde b}
\nc{\tc}{\tilde c}
\nc{\te}{\tilde e}
\nc{\tf}{\widetilde f}
\nc{\tg}{\widetilde g}
\nc{\ti}{\tilde i}
\nc{\tj}{\tilde j}
\nc{\tk}{\widetilde k}
\nc{\tl}{\tilde l}
\nc{\tm}{\widetilde m}
\nc{\tn}{\tilde n}
\nc{\tp}{\tilde{p}}
\nc{\tq}{\widetilde{q}}
\nc{\trr}{{\tilde r}}
\nc{\ts}{{\tilde s}}
\nc{\tu}{{\tilde u}}
\nc{\tv}{{\tilde v}}
\nc{\tw}{{\tilde w}}
\nc{\tx}{{\tilde x}}
\nc{\ty}{{\tilde y}}
\nc{\tz}{\tilde z}
\nc{\tA}{{\widetilde A}}
\nc{\tAbar}{{\ol \tA}}
\nc{\tB}{{\widetilde B}}
\nc{\tC}{{\widetilde C}}
\nc{\tD}{{\widetilde D}}
\nc{\tE}{{\widetilde E}}
\nc{\tF}{{\widetilde F}}
\nc{\tG}{{\widetilde G}}
\nc{\tcG}{{\widetilde \cG}}
\nc{\tH}{{\widetilde H}}
\nc{\tI}{{\widetilde I}}
\nc{\tcI}{{\widetilde \cI}}
\nc{\tJ}{{\widetilde J}}
\nc{\tJbar}{{\ol {\tilde J}}}
\nc{\tK}{{\widetilde K}}
\nc{\tL}{{\widetilde L}}
\nc{\tcL}{{\widetilde \cL}}
\nc{\tcLbar}{{\ol \tcL}}
\nc{\tM}{{\widetilde M}}
\nc{\tN}{{\widetilde N}}
\nc{\tcN}{{\widetilde \cN}}
\nc{\tP}{{\widetilde P}}
\nc{\tQ}{{\widetilde Q}}
\nc{\tR}{{\widetilde R}}
\nc{\tS}{\widetilde{S}}
\nc{\tT}{\widetilde{T}}
\nc{\tU}{\widetilde{U}}
\nc{\tUU}{\widetilde{\UU}}
\nc{\tV}{\widetilde{V}}
\nc{\tcV}{\widetilde{\cV}}
\nc{\tW}{\widetilde{W}}
\nc{\tcF}{\widetilde{{\cal F}}}
\nc{\tX}{\widetilde{X}}
\nc{\tY}{\widetilde{Y}}
\nc{\tcZ}{\tilde{\cZ}}
\nc{\tcZbar}{\ol{\tcZ}}

\nc{\ha}{\hat a}
\nc{\hb}{\hat b}
\nc{\hc}{\widehat c}
\nc{\hd}{\widehat d}
\nc{\he}{\widehat e}
\nc{\hf}{\widehat f}
\nc{\hg}{\widehat g}
\nc{\hh}{\widehat h}
\nc{\hm}{\widehat m}
\nc{\hn}{\widehat n}
\nc{\hp}{\widehat p}
\nc{\hq}{\widehat q}
\nc{\hr}{\widehat r}
\nc{\hs}{\widehat s}
\nc{\hv}{\widehat v}
\nc{\hw}{\widehat w}
\nc{\hx}{\widehat x}
\nc{\hy}{\widehat y}
\nc{\hz}{\widehat z}
\nc{\zhat}{\hat z}
\nc{\hA}{\widehat{A}}
\nc{\hB}{\widehat{B}}
\nc{\hC}{\widehat{C}}
\nc{\hD}{\widehat{D}}
\nc{\hE}{\widehat{E}}
\nc{\hF}{\widehat{F}}
\nc{\hcF}{\widehat{\cF}}
\nc{\hG}{\widehat{G}}
\nc{\hcG}{\widehat{\cG}}
\nc{\hH}{\widehat{H}}
\nc{\hI}{\widehat{I}}
\nc{\hcI}{\widehat{\cI}}
\nc{\hJ}{\widehat{J}}
\nc{\hK}{\widehat{K}}
\nc{\hL}{\widehat{L}}
\nc{\hcL}{\widehat{\cL}}
\nc{\hM}{\widehat M}
\nc{\hcM}{\widehat{\cM}}
\nc{\hN}{\widehat{N}}
\nc{\hO}{\widehat{O}}
\nc{\hcO}{\widehat{\cO}}
\nc{\hP}{\widehat{P}}
\nc{\hQ}{\widehat{Q}}
\nc{\hcQ}{\widehat{\cQ}}
\nc{\hcR}{\widehat{\cR}}
\nc{\hR}{\widehat{R}}
\nc{\hS}{\widehat{S}}
\nc{\hcS}{\widehat{\cS}}
\nc{\hT}{\widehat{T}}
\nc{\hU}{\widehat{U}}
\nc{\hV}{\widehat V}
\nc{\hcV}{\widehat \cV}
\nc{\hX}{\widehat X}
\nc{\hcZ}{\widehat \cZ}
\nc{\hcZbar}{\ol{\widehat \cZ}}

\nc{\heta}{\widehat{\eta}}
\nc{\hal}{\widehat \alpha}
\nc{\hbeta}{\widehat \beta}
\nc{\hphi}{\widehat{\phi}}
\nc{\hkap}{\hat{\kappa}}
\nc{\hchi}{\widehat{\chi}}
\nc{\hpsi}{\widehat{\psi}}
\nc{\hgam}{\widehat{\gam}}
\nc{\hPhi}{\hat{\Phi}}
\nc{\hPsi}{\hat{\Psi}}
\nc{\hGam}{\hat{\Gam}}
\nc{\omhat}{\widehat{\om}}
\nc{\htha}{\hat{\tha}}
\nc{\hrho}{\widehat{\rho}}
\nc{\hdel}{\widehat{\del}}

\nc{\w}{\wedge}

\nc{\vb}{\vec b}
\nc{\vc}{\vec c}
\nc{\vd}{\vec d}
\nc{\ve}{\vec e}
\nc{\vf}{\vec f}
\nc{\vg}{\vec g}
\nc{\vh}{\vec h}
\nc{\vp}{\vec p}
\nc{\vq}{\vec q}
\nc{\vr}{\vec r}
\nc{\vs}{\vec s}
\nc{\vv}{\vec v}
\nc{\vw}{\vec w}
\nc{\vx}{\vec x}
\nc{\vy}{\vec y}
\nc{\vz}{\vec z}

\nc{\vB}{\vec B}
\nc{\vC}{\vec C}
\nc{\vD}{\vec D}
\nc{\vE}{\vec E}
\nc{\vF}{\vec F}
\nc{\vG}{\vec G}
\nc{\vH}{\vec H}
\nc{\vP}{\vec P}
\nc{\vQ}{\vec Q}
\nc{\vR}{\vec R}
\nc{\vS}{\vec S}
\nc{\vV}{\vec V}
\nc{\vW}{\vec W}
\nc{\vX}{\vec X}
\nc{\vY}{\vec Y}
\nc{\vZ}{\vec Z}

\nc{\ol}{\overline}
\nc{\abar}{\ol{a}}
\nc{\bbar}{\ol{b}}
\nc{\cbar}{\ol{c}}
\nc{\dbar}{\ol{d}}
\nc{\ebar}{\ol{e}}
\nc{\fbar}{\ol{f}}
\nc{\gbar}{\ol{g}}
\nc{\ibar}{\ol{\imath}}
\nc{\jbar}{\ol{\jmath}}
\nc{\kbar}{\ol{k}}
\nc{\lbar}{\ol{l}}
\nc{\mbar}{\ol{m}}
\nc{\nbar}{\ol{n}}
\nc{\pbar}{\ol{p}}
\nc{\qbar}{\ol{q}}
\nc{\rbar}{\ol{r}}
\nc{\sbar}{\ol{s}}
\nc{\ubar}{\ol{u}}
\nc{\vbar}{\ol{v}}
\nc{\wbar}{\ol{w}}
\nc{\xbar}{\ol{x}}
\nc{\ybar}{\ol{y}}
\nc{\zbar}{\ol{z}}

\nc{\Abar}{\ol{A}}
\nc{\Bbar}{\ol{B}}
\nc{\Cbar}{\ol{C}}
\nc{\Dbar}{\ol{D}}
\nc{\Ebar}{\ol{E}}
\nc{\Fbar}{\ol{F}}
\nc{\Jbar}{\ol{J}}
\nc{\Kbar}{\ol{K}}
\nc{\cKbar}{\ol{\cK}}
\nc{\Lbar}{\ol{L}}
\nc{\cLbar}{\ol{\cL}}
\nc{\Mbar}{\ol{M}}
\nc{\Nbar}{\ol{N}}
\nc{\Pbar}{\ol{P}}
\nc{\Qbar}{\ol{Q}}
\nc{\Rbar}{\ol{R}}
\nc{\Sbar}{\ol{S}}
\nc{\Tbar}{\ol{T}}
\nc{\Ubar}{\ol{U}}
\nc{\Vbar}{\ol{V}}
\nc{\cVbar}{\ol{\cV}}
\nc{\Wbar}{\ol{W}}
\nc{\cWbar}{\ol{\cW}}
\nc{\Xbar}{{\overline X}}
\nc{\Ybar}{{\overline Y}}
\nc{\Zbar}{{\overline Z}}
\nc{\cZbar}{{\overline \cZ}}

\nc{\epsbar}{\ol{\epsilon}}
\nc{\albar}{\ol{\al}}
\nc{\Albar}{\ol{\Al}}
\nc{\betabar}{\ol{\beta}}
\nc{\Betabar}{\ol{\Beta}}
\nc{\lambar}{\ol{\lambda}}
\nc{\kapbar}{\ol{\kappa}}
\nc{\zetabar}{\ol{\zeta}}
\nc{\Zetabar}{\ol{\Zeta}}
\nc{\taubar}{\ol{\tau}}
\nc{\Taubar}{\ol{\Tau}}
\nc{\psibar}{\ol{\psi}}
\nc{\Psibar}{\ol{\Psi}}
\nc{\tpsibar}{\ol{\tpsi}}
\nc{\tPsibar}{\ol{\tPsi}}
\nc{\phibar}{\ol{\phi}}
\nc{\Phibar}{\ol{\Phi}}
\nc{\chibar}{\ol{\chi}}
\nc{\mubar}{\ol{\mu}}
\nc{\nubar}{\ol{\nu}}
\nc{\rhobar}{\ol{\rho}}
\nc{\ombar}{\ol{\om}}
\nc{\Ombar}{\ol{\Om}}
\nc{\Deltabar}{\ol{\Delta}}
\nc{\Thetabar}{\ol{\Theta}}
\nc{\xibar}{\ol{\xi}}
\nc{\Xibar}{\ol{\Xi}}

\nc{\Dthbar}{\ol{\rm D3}}

\nc{\fdot}{\dot{f}}
\nc{\gdot}{\dot{g}}
\nc{\pdot}{\dot{p}}
\nc{\qdot}{\dot{q}}
\nc{\rdot}{\dot{r}}
\nc{\sdot}{\dot{s}}
\nc{\tdot}{\dot{t}}
\nc{\udot}{\dot{u}}
\nc{\vdot}{\dot{v}}
\nc{\wdot}{\dot{w}}
\nc{\xdot}{\dot{x}}
\nc{\xddot}{\ddot{x}}
\nc{\ydot}{\dot{y}}
\nc{\zdot}{\dot{z}}
\nc{\yddot}{\ddot{y}}

\nc{\Adot}{\dot{A}}
\nc{\Bdot}{\dot{B}}
\nc{\Cdot}{\dot{C}}
\nc{\Udot}{\dot{U}}
\nc{\Vdot}{\dot{V}}
\nc{\Wdot}{\dot{W}}

\nc{\taudot}{\dot{\tau}}
\nc{\phidot}{\dot{\phi}}
\nc{\psidot}{\dot{\psi}}
\nc{\chidot}{\dot{\chi}}
\nc{\sinp}{s_{\phi}}
\nc{\cosp}{c_{\phi}}
\nc{\tanp}{t_{\phi}}
\nc{\spone}{s_{\phi_1}}
\nc{\cpone}{c_{\phi_1}}
\nc{\tpone}{t_{\phi_1}}
\nc{\sptwo}{s_{\phi_2}}
\nc{\cptwo}{c_{\phi_2}}
\nc{\tptwo}{t_{\phi_2}}
\nc{\spth}{s_{\phi_3}}
\nc{\cpth}{c_{\phi_3}}
\nc{\tpth}{t_{\phi_3}}
\nc{\calp}{c_{\al}}
\nc{\salp}{s_{\al}}

\nc{\csch}{{\rm csch}}
\nc{\sech}{{\rm sech}}

\nc{\cothzlami}{\coth(z-\lam_i)}
\nc{\coshzlami}{\cosh(z-\lam_i)}
\nc{\sinhzlami}{\sinh(z-\lam_i)}

\nc{\cothzlamj}{\coth(z-\lam_j)}
\nc{\coshzlamj}{\cosh(z-\lam_j)}
\nc{\sinhzlamj}{\sinh(z-\lam_j)}

\nc{\cothlamij}{\coth(\lam_i-\lam_j)}
\nc{\coshlamij}{\cosh(\lam_i-\lam_j)}
\nc{\sinhlamij}{\sinh(\lam_i-\lam_j)}

\nc{\bah}{{\mathbf {\hat{A}}}}
\nc{\bX}{{\mathbf X}}
\nc{\ba}{{\bf a}}
\nc{\bb}{{\bf b}}
\nc{\bc}{{\bf c}}
\nc{\bd}{{\bf d}}
\nc{\bg}{{\bf g}}
\nc{\bk}{{\bf k}}
\nc{\bl}{{\bf l}}
\nc{\bm}{{\bf m}}
\nc{\bn}{{\bf n}}
\nc{\bo}{{\bf o}}
\nc{\bp}{{\bf p}}
\nc{\bq}{{\bf q}}
\nc{\br}{{\bf r}}
\nc{\bs}{{\bf s}}
\nc{\bt}{{\bf t}}
\nc{\bu}{{\bf u}}
\nc{\bv}{{\bf v}}
\nc{\bw}{{\bf w}}
\nc{\bx}{{\bf x}}
\nc{\by}{{\bf y}}
\nc{\bz}{{\bf z}}
\nc{\bom}{{\bf \om}}
\nc{\bombar}{{\mathbf \ombar}}
\nc{\bPhi}{{\bf \Phi}}

\nc{\rma}{{\rm a}}
\nc{\rmb}{{\rm b}}
\nc{\rmc}{{\rm c}}
\nc{\rmd}{{\rm d}}
\nc{\rmg}{{\rm g}}
\nc{\rk}{{\rm k}}
\nc{\rml}{{\rm l}}
\nc{\rmm}{{\rm m}}
\nc{\rmn}{{\rm n}}
\nc{\rmo}{{\rm o}}
\nc{\rmp}{{\rm p}}
\nc{\rmq}{{\rm q}}
\nc{\rmr}{{\rm r}}
\nc{\rms}{{\rm s}}
\nc{\rmt}{{\rm t}}
\nc{\rmu}{{\rm u}}
\nc{\rmv}{{\rm v}}
\nc{\rmw}{{\rm w}}
\nc{\rmx}{{\rm x}}
\nc{\rmy}{{\rm y}}
\nc{\rmz}{{\rm z}}

\nc{\dal}{\dot{\al}}
\nc{\thadot}{\dot{\tha}}
\nc{\thab}{\bar{\theta}}
\nc{\thal}{\theta^{\al}}
\nc{\thdal}{\bar{\theta}^{\dal}}

\nc{\thsigthm}{\tha \sigma^m \thab}
\nc{\thsigthn}{\tha \sigma^n \thab}

\nc{\Dal}{D_{\al}}
\nc{\Ddal}{\bar{D}_{\dal}}
\nc{\CDal}{{\cal D}_{\al}}
\nc{\CDdal}{\bar{\cal D}_{\dal}}

\nc{\eq}[1]{{(\ref{#1})}}
\nc{\eqtwo}[2]{{(\ref{#1},\ref{#2})}}
\nc{\eqthree}[3]{(\ref{#1},\ref{#2},\ref{#3})}
\nc{\eqfour}[4]{(\ref{#1},\ref{#2},\ref{#3},\ref{#4})}
\nc{\eqfive}[5]{(\ref{#1},\ref{#2},\ref{#3},\ref{#4,\ref{#5}})}
\nc{\non}{\nonumber}
\nc{\Fzero}{F_{(0)}}
\nc{\Ftwo}{F_{(2)}}
\nc{\Ffour}{F_{(4)}}
\nc{\Fone}{F_{(1)}}
\nc{\Fthree}{F_{(3)}}
\nc{\Ffive}{F_{(5)}}
\nc{\Fn}{F_{(n)}}
\nc{\Fp}{F_{(p)}}

\nc{\tFzero}{\tF_{(0)}}
\nc{\tFtwo}{\tF_{(2)}}
\nc{\tFfour}{\tF_{(4)}}
\nc{\tFone}{\tF_{(1)}}
\nc{\tFthree}{\tF_{(3)}}
\nc{\tFfive}{\tF_{(5)}}
\nc{\tFn}{\tF_{(n)}}
\nc{\tFp}{\tF_{(p)}}

\nc{\Czero}{C_{(0)}}
\nc{\Ctwo}{C_{(2)}}
\nc{\Cfour}{C_{(4)}}
\nc{\Cone}{C_{(1)}}
\nc{\Cthree}{C_{(3)}}
\nc{\Cfive}{C_{(5)}}
\nc{\Cn}{C_{(n)}}

\nc{\equ}{{\rm eq}}
\def\Im{{\rm Im \hspace{0.5mm} }}

\nc{\vol}{{\rm vol}}
\nc{\Ainf}{A_{\infty}}
\nc{\End}{{\rm End}}
\nc{\Ext}{{\rm Ext}}
\nc{\IIB}{{\rm IIB}}
\nc{\Ad}{{\rm Ad}}
\nc{\IIA}{{\rm IIA}}
\nc{\AdS}{{\rm AdS}}
\nc{\CFT}{{\rm CFT}}
\nc{\diag}{{\rm diag}}
\nc{\Log}{{\rm Log}}
\nc{\Dslash}{\ensuremath \raisebox{0.025cm}{\slash}\hspace{-0.32cm} D}
\nc{\cDslash}{\ensuremath \raisebox{0.025cm}{\slash}\hspace{-0.32cm} \cD}
\nc{\omslash}{\om\!\!\!/}
\nc{\no}{\!:\!\!}
\nc{\ointdz}{\oint\frac{dz}{2\pi i}}
\nc{\ointdzone}{\oint\frac{dz_1}{2\pi i}}
\nc{\ointdztwo}{\oint\frac{dz_2}{2\pi i}}
\nc{\ointdzb}{\oint\frac{d\zbar}{2\pi i}}
\nc{\ointdzbone}{\oint\frac{d\zbar_1}{2\pi i}}
\nc{\ointdzbtwo}{\oint\frac{d\zbar_2}{2\pi i}}
\nc{\dz}{\frac{dz}{2\pi i}}
\nc{\dzb}{\frac{d\zbar}{2\pi i}}
\nc{\bpm}{\begin{pmatrix}}
\nc{\epm}{\end{pmatrix}}
 \nc{\bitem}{\begin{itemize}}
 \nc{\eitem}{\end{itemize}}
 \nc{\exercise}{\vskip 2mm \noindent {\bf Exercise:}}
 \nc{\definition}{\vskip 2mm \noindent {\bf Definition:}}
 
\begin{document}

\vspace{0.5cm}
\begin{center}
\baselineskip=13pt {\LARGE \bf{AdS Black Holes from Duality   \\ \vskip 4mm in Gauged Supergravity }}
 \vskip1.5cm 
Nick Halmagyi and Thomas Vanel\\ 
\vskip0.5cm
\textit{Laboratoire de Physique Th\'eorique et Hautes Energies,\\
Universit\'e Pierre et Marie Curie, CNRS UMR 7589, \\
F-75252 Paris Cedex 05, France}\\
\vskip0.5cm
halmagyi@lpthe.jussieu.fr \\ 
vanel@lpthe.jussieu.fr \\ 
\end{center}

\begin{abstract}
We study and utilize duality transformations in a particular STU-model of four dimensional gauged supergravity. This model is a truncation of the de Wit-Nicolai $\cN=8$ theory and as such has a lift to eleven-dimensional supergravity on the seven-sphere. Our duality group is $U(1)^3$ and while it can be applied to any solution of this theory, we consider known asymptotically AdS$_4$, supersymmetric black holes and focus on duality transformations which preserve supersymmetry. For static black holes we generalize the supersymmetric solutions of Cacciatori and Klemm from three magnetic charges to include two additional electric charges and argue that this is co-dimension one in the full space of supersymmetric static black holes in the STU-model. These new static black holes have nontrivial profiles for axions.
 For rotating black holes, we generalize the known two-parameter supersymmetric solution to include an additional parameter which represents scalar hair. When lifted to M-theory, these black holes correspond to the near horizon geometry of a stack of BPS rotating M2-branes, spinning on an $S^7$ which is fibered non-trivially over a Riemann surface. 
\end{abstract} 

\newpage
\section{Introduction}

Black branes in gauged supergravity are of particular interest due to their ability to possess AdS asymptotics and they have numerous applications to holography.  Somewhat recently \cite{Cacciatori:2009iz} an exact analytic solution for static quarter-BPS black holes was found as well as an analytic quarter-BPS rotating black hole in \cite{Klemm:2011xw}. This work was performed in an  $\cN=2$ truncation of the four dimensional $\cN=8$ gauged supergravity theory of de Wit-Nicolai \cite{deWit:1982ig}  and as such these black holes can be lifted to M-theory. Generalizing these solutions to new analytic families of supersymmetric AdS$_4$ black holes is the focus of our current work.

The static black holes of \cite{Cacciatori:2009iz} can be understood within the context of the far-reaching work of Maldacena and Nunez \cite{Maldacena:2000mw}; in M-theory they correspond to a stack of M2-branes wrapped on a Riemann surface $\Sig_g$ of genus $g\geq 0$. The initial work \cite{Maldacena:2000mw} found AdS$_p\times \Sig_g$ geometries in $(p+2)$-dimensional gauged supergravity only when $g>1$ and $p=1,3$ but the method was clearly universal and there has since been much work establishing the phase space of solutions for arbitrary genus and various $p$%
\footnote{See for example \cite{Nunez:2001pt, Gauntlett:2001qs, Cacciatori:2009iz, Cucu:2003bm, Cucu:2003yk, Bah:2012dg, Almuhairi:2011ws, Benini:2013cda} and some aspects are nicely reviews in \cite{Gauntlett:2003di}.}. 
The work of CK should be singled out for special mention since this is the only example with non-trivial scalar field profiles where the entire black-brane geometry is known analytically%
\footnote{We should mention the constant scalar black branes which exist for $p=2,3$ and $g>1$ \cite{Caldarelli1999, Klemm:2000nj}.}. In addition, from a purely general relativistic point of view, four dimensional black holes with spherical horizons are traditionally of substantial interest as compared to black branes in higher dimensions.

In this work we apply a tried and true method of generating solutions in supergravity theories: the awful power of the Geroch group \cite{Geroch:1970nt}. In section \ref{sec:STU} we find by explicit computation that the bosonic sector of our gauged STU model has a $G=U(1)^3$ invariance and one can use this group to act on any solution of the theory. We denote the diagonal $U(1)$ subgroup of $G$ by $U(1)_g$ and find reason to conjecture that $G/U(1)_g$ is in addition a symmetry of the fermionic sector of the theory.

In section \ref{sec:GenerateStatic} we look at the CK solutions. They depend on three charges; there are initially four charges but one BPS condition enforces a Dirac quantization condition and reduces this to a three dimensional parameter space. We act on the CK solutions with the two generators of $G/U(1)_g$ and generate static BPS black holes with two additional charges. In the symplectic frame adapted to the M-theory lift, the CK solution has purely magnetic charges whereas our two additional parameters are electric charges. Another point of comparison is that our new solutions have non-trivial axions whereas in the CK solutions the axions are trivial. Acting on the CK solutions with $U(1)_g\subset G$ breaks the supersymmetry of the solutions and also appears to violate the Dirac quantization condition, as a result we focus on the generators of $G/U(1)_g$. We also act on the CK solutions with equal magnetic charges and generate a new parameter corresponding to scalar hair.

In section \ref{sec:GenerateRotation} we perform a similar action of $G/U(1)_g$ on the BPS rotating black holes of \cite{Klemm:2011xw}. The solutions of \cite{Klemm:2011xw} have equal magnetic charges which are inversely proportional to the gauge coupling and they depend on two parameters. One parameter corresponds to angular momentum the other represents a deformation of the boundary M2-brane theory. The static limit is a solution from \cite{Cacciatori:2009iz} with a single parameter corresponding to a deformation of the boundary M2-brane theory. Another limit sets the deformation parameter to zero and corresponds to the constant scalar black hole with rotation. While this constant scalar black hole is a fixed point of our duality group, from the solutions of \cite{Klemm:2011xw} we generate one additional parameter for scalar hair. The full solution space of BPS rotating black holes now has three parameters; angular momentum, one deformation parameter and one parameter of scalar hair.

When lifted to M-theory the charges of the CK solutions correspond to twists of the $S^7$ bundle over $\Sig_g$ \cite{Maldacena:2000mw}. From another point of view one can view these solutions as the near horizon limit of a stack of M2-branes wrapping a Riemann surface inside a local Calabi-Yau fivefold $X_5$ which is the product of four line bundle over $\Sig_g$. The magnetic charges of the CK solution are proportional to the Chern numbers of these four line bundles. In this same duality frame, the electric charges we find correspond to the spin of the M2-branes along a pair of circles: $U(1)^2\subset S^7$. 

\section{STU-model of gauged supergravity from M-theory} \label{sec:STU}

We start in the symplectic duality frame where the STU-model of four dimensional supergravity has the prepotential
\be\label{FSTU}
F=-\frac{X^1X^2 X^3}{X^0}\,.
\ee
Using the notation of appendix \ref{app:SpGeometry} this implies that $d_{123}=\frac{1}{6}$ and $\hd^{123}=\frac{32}{3}$. This model has the vector-multiplet scalar manifold
\be
\cM_v = \Blp \frac{SL(2,\RR)}{U(1)}\Brp^3
\ee
and thus the global symmetry $\bslb SL(2,\RR)\bsrb^3$.
We include a very specific dyonic gauging, namely we take
\be
\cG=\bpm g^\Lam \\ g_{\Lam} \epm\,,\quad\quad g^{\Lam}=\bpm 0 \\ g^1 \\g^2 \\g^3\epm\,,\quad\quad g_\Lam = \bpm g_0 \\0\\0\\0 \epm
\ee
and using a duality symmetry from appendix \ref{app:Duality} with 
\bea
&& \beta=\log\Bslb \frac{g_0}{g}\Bsrb\,,\ \quad\quad B^i_{\ i}=-\log\Bslb-\frac{g^i(g_0)^{1/3}}{g^{4/3}}\Bsrb\,, \non \\
&& a^i=b_j=0\,,\quad\quad\quad  B^{i}_{\ j}=0\,,\quad {\rm for }\ \ i\neq j
\eea
we set the magnitudes of the gauge couplings equal
\be\label{STUgaugings}
g^{\Lam}=-\bpm 0 \\ g \\g \\g\epm\,,\quad\quad g_\Lam = \bpm g \\0\\0\\0 \epm\,.
\ee

There is a simple reason for choosing this seemingly obscure gauging: this model is known to be a truncation of $\cN=8$, de Wit-Nicolai theory \cite{deWit:1981eq, Duff:1999gh, Cvetic1999b} with $n_v=3$ and can thus be uplifted to M-theory%
\footnote{There has been recent work \cite{Nicolai:2011cy} refining the explicit uplift \cite{deWit:1986iy} of this $\cN=8$ theory to eleven dimensional supergravity and thus proving that it is a consistent truncation.}. The model given by \eq{FSTU} and \eq{STUgaugings} is related by a symplectic transformation 
\be\label{Sdefinition}
\cS=\bpm A & B\\ C&D\epm\,,\quad\quad A=D=\diag\{1,0,0,0\}\,,\quad B=-C=\diag\{0,1,1,1\}
\ee
to the perhaps more familiar model with prepotential, gaugings and sections given by
\bea
\bvF&=&-2i\sqrt{\bvX^0 \bvX^1 \bvX^2 \bvX^3}\,,\quad \bvg^\Lam=0\,,\quad\quad \bvg_\Lam=g\,, \label{SympFrame1}\\
 \bvX^\Lam&=&\bpm1\\ -z^2 z^3\\-z^3 z^1\\-z^1 z^2\epm \,,\quad 
 \bvF_\Lam= \bpm z^1z^2 z^3 \\ -z^1\\-z^2 \\ -z^3\epm \label{SympFrame2}
\eea
but we are particularly fond of the frame \eq{FSTU} because it makes the action of the symplectic group $Sp(2n_v+2,\RR)$ manifest and thus is the natural frame to understand the unbroken symmetries. Of course both frames are physically indistinguishable.

With dyonic gaugings such as \eq{STUgaugings} it is convenient to use the formalism of \cite{Dall'Agata:2010gj} which is a natural symplectic completion of the electrically gauged theory. In particular the scalar potential is
\bea\label{scalarPot1}
V_g&=& g^{i\jbar} D_i \cL D_{\jbar} \cLbar - 3 |\cL|^2
\eea
where we have defined the symplectic invariant quantities 
\be
\cL=\langle  \cG,\cV \rangle\,,\quad \quad \cL_i=\langle  \cG,D_i\cV \rangle
\ee
and $\langle .,.\rangle$ is the symplectic product of two symplectic vectors. For the STU model with gaugings given by \eq{STUgaugings}, the scalar potential has the following explicit form:
\be\label{scalarPot2}
V_g=-g^2\sum_{i=1}^3 \Bslb \frac{1}{y^i}+y^i+ \frac{(x^i)^2}{y^i} \Bsrb\,.
\ee
Our first goal is to analyze the subgroup of $\bslb SL(2,\RR)\bsrb^3$ which remains unbroken in the bosonic sector of the gauged theory to do so it is sufficient to analyze the invariances of $V_g$. 

\subsection{The basics of $SL(2,\RR)/U(1)$}\label{sec:SL2R}
This section contains some details about the coset $SL(2,\RR)/U(1)$. We are aware that this material is quite elementary but see no reason not to spell out our steps in modest detail. Indeed, the symmetries of this particularly interesting STU-model of gauged supergravity are remarkably straightforward, nonetheless to the best of our knowledge have never been worked out or utilized. 

The coset representative is
\be\label{Vhatdef}
V= e^{H \frac{\phi}{2}} e^{E\, \chi}
\ee
where the generators of $\sl(2,\RR)$ are
\be
H=\bpm 1 & 0 \\ 0 & -1  \epm \,,\quad E=\bpm 0 & 1 \\ 0 & 0 \epm\,,\quad F=\bpm 0 & 0 \\1 & 0 \epm\,.
\ee
To construct the metric on the coset, one takes 
\be\label{MCosetdef}
M=V^T V
\ee
and under the right action of $\Lam\in SL(2,\RR)$ these transform as
\bea
V&\ra&V\Lam\,,\quad\quad M\ra \Lam^T M \Lam \label{Vtrans}\,.
\eea
The transformation \eq{Vtrans} ruins the parametrization \eq{Vhatdef} but one uses a compensating, local, left acting $SO(2)$ transformation to bring $V$ back to the form \eq{Vhatdef}.
From \eq{Vtrans} we see that $\Tr M$ is invariant under $\Lam \in SO(2)$. The kinetic terms for the coset are then given by
\be
\cL_{kin}= -\frac{1}{4} \Tr (\del_\mu M \del^\mu M^{-1})
\ee
and are invariant under \eq{Vtrans} for $\Lam \in SL(2,\RR)$.

Explicitly, using \eq{Vhatdef} and \eq{MCosetdef} we have
\be
M=\bpm e^\phi &  e^\phi \chi\\e^\phi \chi &  e^{-\phi}+ e^\phi \chi^2 \epm
\ee
and using the standard co-ordinate redefinition
\be
z=x+iy = \chi+i e^{-\phi}
\ee
we find that 
\be
\Tr M = \frac{1}{y} + y +\frac{x^2}{y}\,.
\ee
So we see that the scalar potential of our gauged supergravity theory \eq{scalarPot2} is given by canonical objects from the coset:
\be
V_g =- g^2 \sum_{i=1}^3 \Tr M_i
\ee
where $M_i$ is \eq{MCosetdef} for the $i$-th $SL(2,\RR)/U(1)$ coset. Thus we have demonstrated that the scalar potential and thus the bosonic sector of the STU model of section \ref{sec:STU} is invariant under
\be\label{SO2cubed}
SO(2)^3 \subset  SL(2,\RR)^3\,.
\ee

\subsection{Embedding $SO(2)^3$ into $Sp(2n_v+2,\RR)$}

We now embed this symmetry group $SO(2)^3$ into $Sp(8,\RR)$ using the work of \cite{deWit:1991nm, deWit:1992wf}, key aspects of this work are summarized in appendix \ref{app:Duality}. The three rotations corresponding to \eq{SO2cubed} are given by the exponentiation of the elements $\underline{\cS}\in \sp(8,\RR)$ from \eq{Sunderline} with
\be
\beta=B^{i}_{\ j}=0\,,\quad\quad a^i=-b_i\,.
\ee
We find that these are given by
\bea\label{Oitheta}
\cO_i(\al)&=& \bpm Q_i(\al) & R_i(\al) \\ S_i(\al) & T_i(\al)\epm
\eea
where
\bea
Q_1(\al)&=&T_1(\al)= \bpm 
c_\al  & s_\al & 0 &0\\
-s_\al & c_\al & 0 & 0\\
0 & 0 &c_\al  & 0\\
0& 0 & 0 &c_\al
\epm\,,\quad
R_1(\al) =-S_1(\al)= \bpm 0 & 0&0&0 \\
0 & 0&0&0 \\
0 & 0& 0 & -s_\al\\
0 & 0& -s_\al&0
 \epm\,, \non \\
Q_2(\al)&=&T_2(\al)= \bpm 
c_\al  & 0 & s_\al &0\\
0 & c_\al & 0 & 0\\
-s_\al& 0 &c_\al  & 0\\
0& 0 & 0 &c_\al
\epm\,,\quad
R_2(\al) =-S_2(\al)= \bpm 0 & 0&0&0 \\
0 & 0&0&-s_\al\\
0 & 0&0&0\\
0 & -s_\al&0&0 \epm\,, \non \\
Q_3(\al)&=&T_3(\al)= \bpm 
c_\al  & 0 & 0 &s_\al\\
0 & c_\al & 0 & 0\\
0 & 0 &c_\al  & 0\\
-s_\al& 0 & 0 &c_\al
\epm\,,\quad
R_3 (\al) =-S_3(\al)= \bpm 0 & 0&0&0 \\
0 & 0&-s_\al&0\\
0 & -s_\al&0&0\\
0 & 0&0&0 \epm \non
\eea
and we use the notation $s_\al=\sin \al$ and $c_\al=\cos \al$.

We know from section \ref{sec:SL2R} that simultaneously acting with $\cO_i$ on both the sections $\cV$ and the vector fields is a symmetry of the Lagrangian. Now by construction the theory is invariant under the simultaneous action of any symplectic matrix $\cT$ on the gaugings $\cG$, charges $\cQ$ and the sections $\cV$:
\be
(\cG,\cQ,\cV)\ra (\cT\cG,\cT\cQ,\cT\cV)\,,\ \ \ \ \cT\in Sp(2n_v+2,\RR)
\ee
and so we can surmise that for our particular theory we could equally well just act on the gaugings 
\be 
\cG\ra \cO_i(\al) \cG
\ee
and this should be a symmetry of the Lagrangian. Indeed explicit calculation shows this to be true.

\subsection{Two simple generators}

For two of these transformations we can see this quite explicitly since for the particular gaugings \eq{STUgaugings} something even stronger is true, the gaugings themselves are invariant:
\bea \label{SimpleOTrans}
\cO_{12}(\al) \cG&=&\cG\,,\quad\quad\quad
\cO_{23}(\al)  \cG=\cG
\eea
where
\be
\cO_{ij}(\al)=\cO_i(\al)\cO_j^{-1}(\al)\,.
\ee
This leads us to conclude that the generators $\cO_{12}(\al)$ and $\cO_{23}(\al)$ commute with the gauge group. In particular this means that solutions generated using $\cO_{12}$ and $\cO_{23}$ from a supersymmetric seed solution will preserve the same amount of supersymmetry. 

\subsection{The third generator} \label{sec:thirdgen}

The final generator can be taken to be
\be\label{O123def}
\cO_{g}(\al)= \cO_1(\al/3)\cO_2(\al/3)\cO_3(\al/3)
\ee
and we find that the gaugings are not invariant:
\be\label{gaugetransform}
g^\Lam \ra -g\bpm s_\al \\ c_\al \\ c_\al \\ c_\al \epm\,,\quad\quad  g_\Lam \ra -g\bpm -c_\al \\ s_\al \\ s_\al \\s_\al\epm\,.
\ee
Nonetheless the whole bosonic Lagrangian is invariant; the kinetic terms are invariant because this transformation is a duality transformation of the underlying ungauged supergravity theory and we have shown explicitly that the scalar potential is invariant. Note however that the two terms in \eq{scalarPot1} are not separately invariant, only the sum is. As a result we can freely generate solutions to the bosonic equations using $\cO_{123}$.

In \cite{Dall'Agata:2010gj} a comment was made regarding a particular $SO(2)\subset SL(2,\RR)^3$ which is identified with the gauging of the graviphoton and thus what we referred to in the introduction as $U(1)_g$. We understand this generator to be $\cO_{g}$. In fact we find it difficult to make the Dirac quantization condition \eq{DiracQu} compatible with this generator, it is the generators $\cO_{12}$ and $\cO_{23}$ which are particularly useful for our purposes. In a different context \cite{Corrado:2002wx}, it was emphasized to great utility that the duality group of a gauged theory is the commutant of the gauge group inside the duality group of the ungauged theory. In our particular example we understand that the gauge group is identified with the $SO(2)$ generated by%
\footnote{One should note however that before gauging, the scalar fields are neutral under the global $U(1)$ which is gauged. In the gauged theory the scalars are not minimally coupled to any gauge fields.}
 $\cO_{g}$ and the commutant of the gauge group to be the $SO(2)^2$ generated by $\cO_{12}$ and $\cO_{23}$. Solutions generated with $\cO_{g}$ will typically break the supersymmetry of the seed solution and $\cO_{g}$ will not appear in the following sections.

\section{BPS static black holes}\label{sec:GenerateStatic}

We now analyze the action of $\cO_i(\al)$ on the supersymmetric static black holes of \cite{Cacciatori:2009iz}, which we will first review. The metric ansatz is
\bea
ds_{BH}^2&=& -e^{2U} dt^2 + e^{-2U} dr^2 + e^{2(V-U)} d\Sig_g^2
\eea
where $d\Sig_g^2$ is the constant curvature metric on $(S^2,\RR^2,\HH^2)$
and the scalar fields depend only on the radial co-ordinate. The BPS equations can be found in \cite{Dall'Agata:2010gj} but we will not utilize them here. It is however worth mentioning in general there is a Dirac quantization condition $\langle \cG,\cQ\rangle\in \ZZ$ which for supersymmetric solutions is strengthened to 
\be\label{DiracQu}
\langle \cG,\cQ\rangle= -\kappa\,,
\ee
where $\kappa=(1,0,-1)$ for $\Sig_g=(S^2,\RR^2,\HH^2)$ respectively.

\subsection{The supersymmetric static black holes}

The black holes of \cite{Cacciatori:2009iz} require the charges
\be\label{CKCharges}
\cQ=\bpm p^\Lam \\ q_{\Lam}\epm\,,\quad \quad p^\Lam =\bpm p^0 \\0\\0\\0 \epm\,, \quad\quad q_\Lam= \bpm 0\\ q_1 \\q_2 \\q_3 \epm
\ee
and we define some rescaled sections
\be\label{rescaledLM}
\tL^\Lam=e^{V-U} L^\Lam\,,\quad\quad \tM_\Lam = e^{V-U} M_\Lam\,. 
\ee
In the duality frame given by \eq{SympFrame2} the charges would be purely magnetic:
\be
(\bvp^\Lam)^T=(p^0,q_1,q_2,q_3)\,,\quad \quad \brvq_\Lam=0\,.
\ee
The solution is mildly cumbersome but completely explicit, it has recently been extended in \cite{Gnecchi:2013mta} to a large class of $\cN=2$ U(1)-gauged supergravity theories and a covariant form of the solution is presented there. It is given by
\bea
e^{V}&=& \frac{r^2}{R} -v_0\,,\label{Vansatz}\\
\tL^0&=& \frac{r}{4 g R} + \beta^0\,, \label{L0ansatz}\\
\tM_i&=& \frac{r}{4 g R} + \beta_i\,,\label{Miansatz}
\eea
where $R$ is the AdS$_4$ radius 
\be
R=\frac{1}{\sqrt{2}\, g}
\ee
and%
\footnote{To maintain covariance in the expression for $v_0$ we have left $g^i$ which should be set $g^i=-g$.} 
\bea
\beta^0 &=& \frac{\eps}{2\sqrt{2}\,g}\sqrt{ \frac{v_0}{2R}-gp^0 }\,, \label{beta0sol}\\
\beta_i&=& -\frac{\eps}{2\sqrt{2}\, g} \sqrt{\frac{v_0}{2R}- g q_i}\,,  \label{betaisol} \\
v_0&=&2R\Bslb g p^0 +\frac{ 27( d_{ijk} g^i \Pi^j \Pi^k)^2}{32 d_\Pi} \Bsrb\,,
\label{v0sol}
\eea
where $\eps=\pm1$ and $\Pi^i$ is a certain function of the charges:
\bea
\Pi^i&=&- \frac{4}{3g} (2 q_i+p^0-q_1-q_2-q_3)\,. 
\eea
From these expressions one obtains the other metric function $e^U$ and the scalars $y^i$ from \eq{Vansatz}-\eq{Miansatz} and \eq{beta0sol}-\eq{v0sol}:
\bea
e^{4U} &=& \frac{1}{64}\frac{e^{4V}}{\tL^0\tM_1 \tM_2 \tM_3}\,, \quad\quad\quad
y^i= \frac{3}{64}\frac{\hd^{ijk} \tM_j \tM_k}{\sqrt{\tL^0\tM_1 \tM_2 \tM_3}}\,,\quad i=1,2,3\,.
\eea

This CK solution has vanishing axions and is specified by three independent charges; there are four charges \eq{CKCharges} with one constraint \eq{DiracQu}. One would typically not refer to the CK solutions as dyonic since in the symplectic frame \eq{SympFrame2} the gaugings are electric and the charges are purely magnetic. There are regular CK black holes for horizons $\Sig_g$ for all $g\geq 0$ but still regularity places bounds on the values of the magnetic charges.

\subsubsection{Equal charges}\label{sec:equalcharges}
When the charges are all equal then from the above analysis we arrive at the well known flow with constant scalar fields for which $\kappa=-1$ as well as
\be
\Pi^i=0\,,\quad\quad v_0=2R g p\,,\quad\quad\beta^0=\beta_i=0\,.
\ee
Taking into account the Dirac quantization condition \eq{DiracQu} the charges are fixed (they do not give an independent  parameter)
\be\label{chargesequal}
p^0=q_i=\frac{1}{4g}
\ee
and the horizon is at
\be
r=r_h=\frac{R}{\sqrt{2}}
\ee
which is positive and thus the black hole is regular.

There is a whole family of solutions which satisfy \eq{chargesequal} and are missed by the above analysis because of some degeneracy in the BPS equations. This solution has a free parameter $\beta$ corresponding to scalar hair. The metric and sections have
\bea
&& v_0= \frac{R}{2}+16R g^2 \beta^2\,, \\
&& \beta^0=\beta_1=\beta \,, \\ 
&& \beta_2=\beta_3=-\beta
\eea
and the resulting scalar fields are purely imaginary (the axions vanish)
\be
z^1=i\, \frac{r+ \Delta}{r -\Delta}\,,\quad\quad z^2=z^3=i\,,
\ee
where with a view towards the next section we have defined a new parameter
\be
\Delta=4 g R \beta=2\sqrt{2}\,\beta\,.
\ee
This solution was originally found in \cite{Cacciatori:2009iz} from the model with $\hF=-i\hX^0 \hX^1$ and we elaborate in the next section on how this is related to the STU model. This gives the metric
\bea\label{equalchargemetric}
ds_{BH}^2&=& -\frac{\blp r^2-\frac{R^2}{2} -\Delta^2\brp^2}{R^2(r^2-\Delta^2)}dt^2 +\frac{R^2(r^2-\Delta^2)}{\blp r^2-\frac{R^2}{2} -\Delta^2\brp^2} dr^2 + \blp r^2-\Delta^2\brp d\Sig_g^2
\eea
where the metric on $\Sig_g=\HH^2/\Gamma$ is
\bea
d\Sig_g^2&=& d\tha^2 + \sinh^2 \tha d\phi^2\,.
\eea
The horizon is at 
\be
r_h= \sqrt{\frac{R^2}{2}+\Delta^2} \,,
\ee
while the scalar field $z^1$ is singular when 
\be
r=r_s\equiv \Delta\,.
\ee
but $r_h>r_s$ so the singularity is cloaked by a horizon and the black hole is regular. The conserved charges are independent of $\Delta$ but the metric and scalar field  depend  nontrivially on $\Delta$. The $\Delta\ra0$ limit gives the constant scalar black hole. 

The UV behaviour of the $\Delta$ dependence scales as $\cO(\frac{1}{r})$ and in principle there is a choice of quantization schemes \cite{Klebanov:1999tb} which allows us to interpret this as a source {\it or} a vev in the boundary M2-brane theory. To clarify this it is instructive to study the horizon geometry. We find the radius of the horizon to be independent of $\Delta$
\be
R^2_{\Sig_g}=\frac{R^2}{2}
\ee
which is comforting since the Bekenstein-Hawking entropy  should not depend on continuous parameters. However the $AdS_2$ radius does depend on $\Delta$:
\be
R_{AdS_2}^2=\frac{R^2}{4\blp 1+\frac{2\Delta^2}{R^2}\brp} \,.
\ee
By general principles of holography the effective AdS$_2$ radius is a measure of the degrees of freedom in boundary superconformal quantum mechanics. This should not depend on the expectation value of any operator and as such we interpret the $\Delta$ dependence to represent an explicit deformation of the boundary M2-brane theory by a dimension one operator. This is on top of the mass terms induced from the curvature of $\Sig_g$ when twisting of the world-volume M2-brane theory \cite{Maldacena:2000mw}.

\subsection{Duality transformations on the CK black holes}

Our new solutions with non-trivial axions and genuinely dyonic charges are given by 
\bea
e^V&=& e^V|_{CK} \non \\
e^U&=& e^U|_{CK} \non \\
\cV_\al &=& \cO_{12}(\al_1) \cO_{23} (\al_2) \cV_{CK} \label{RotatedSections}\\
\cQ_\al&=&  \cO_{12}(\al_1) \cO_{23} (\al_2) \cQ_{CK} \non \\
\cG_\al&=& \cG \,,\non
\eea
where $\cQ_{CK}$ refers to \eq{CKCharges} and $\cG$ refers to \eq{STUgaugings}.
The scalar fields transform by fractional linear transformations:
\bea
z^1_\al&=&\frac{c_{\al_1} z^1 -s_{\al_1} }{s_{\al_1}z^1+c_{\al_1}}\,, \label{zsol1} \\
z^2_\al&=&\frac{c_{\al_{21}} z^2 -s_{\al_{21}} }{s_{\al_{21}}z^2+c_{\al_{21}}}\,, \label{zsol2} \\
z^3_\al&=&\frac{c_{\al_2} z^3 +s_{\al_2} }{-s_{\al_2}z^3+c_{\al_2}}\,, \label{zsol3}
\eea
where $\al_{21}=\al_2-\al_1$ and one can observe that non-trivial axions are generated. Importantly, one can check that the Dirac quantization condition is invariant:
\be
\langle \cG , \cO_{12}(\al_1) \cO_{23} (\al_2) \cQ_{CK}\rangle =\langle \cG ,  \cQ_{CK}\rangle\,.
\ee

This space of supersymmetric static black holes now depends on five charges; three initial charges from the CK solutions and the parameters $(\al_1,\al_2)$ generate two new charges. As such there is no duality frame where the charges of the entire family are purely magnetic; they are genuinely dyonic black holes. 
In \cite{Halmagyi:2013qoa} a complete solution was found for BPS horizon geometries of the form AdS$_2\times \Sig_g$ in FI-gauged supergravity. It was found in \cite{Halmagyi:2013qoa} that the space of BPS horizon geometries should be $2n_v$-dimensional. The counting works as follows: the gaugings $\cG$ define the theory and therefore are fixed. There are $n_v+1$ electric charges and $n_v+1$ magnetic charges. Then there is the Dirac quantization condition \eq{DiracQu} and in \cite{Halmagyi:2013qoa} one additional constraint was found leaving $2n_v$ parameters. For the model at hand $n_v=3$ and this space is six dimensional. Assuming that every BPS solution of the form AdS$_2\times \Sig_g$ can be completed in the UV to a genuine AdS$_4$ black hole, it would seem there is still one dimension of the black hole solution space missing. We will comment on this further in the conclusions.

For equal charge solutions with \eq{chargesequal}, there is an additional branch of solutions. The charges are invariant under \eq{RotatedSections} but with $\Delta\neq0$ the scalar fields $(z^2,z^3)$ are invariant while $z^1$ transforms according to \eq{zsol1}:
\bea
z^1_\al&=&\frac{2r\Delta s_{2\al}+i (r^2-\Delta^2)}{r^2 +\Delta^2-2r\Delta c_{2\al}} \,, \label{z1static}\\
z^2_\al&=& z^3_\al\ =\ i\,.
\eea
The metric is invariant and given by \eq{equalchargemetric}.
When $\Delta=0$ the whole solution is invariant. The regularity of the black hole can be easily analyzed, when $\al=0$ the scalar $z^1$ diverges at $r=\Delta$ while for $\al\neq 0$ the imaginary part $\Im(z^1)$ vanishes at $r=\Delta$. Nonetheless this is still shielded by the horizon whose position is independent of $\al$. So for the fixed charges \eq{chargesequal} the full solution space is now a family of solutions with two parameters $(\Delta,\al)$. Since the metric does not depend on $\al$ the effective AdS$_2$ radius does not depend on $\al$ and we interpret this mode as an expectation value. One could refer to this as {\it scalar hair}.

\section{Rotating black holes}\label{sec:GenerateRotation}

We now apply our duality transformations to rotating black holes. We focus on the BPS rotating black holes in AdS$_4$ are those of \cite{Klemm:2011xw}, these solutions were originally found in the gauged supergravity model with prepotential and sections given by%
\footnote{To be clear, the hatted variables refer to the model of \eq{FX0X1}, the variables with a breve $``\ \breve{}\ "$ refer to the STU-model in the frame given by \eq{SympFrame1} and \eq{SympFrame2} while the un-hatted, un-breved variables refer to STU model obtained from the cubic prepotential \eq{FSTU}. The duality rotations \eq{Oitheta} act in the frame of \eq{FSTU}.}
\be\label{FX0X1}
\hF=-i \hX^0 \hX^1\,,\quad\quad \hX^\Lam= \bpm 1 \\ \tau\epm\,, \quad\quad \hF_\Lam = \bpm -i\tau \\ -i\epm \,,\quad\quad \tau=x+iy\,.
\ee
This model does not have a frame where it is given by a cubic prepotential but one can embed it into the STU-model in the frame \eq{SympFrame1} and \eq{SympFrame2}. We now describe this embedding in some detail and then the resulting action of the duality group. To do so we take the scalar fields
\bea
&&z^1=i\tau\,, \label{ziembed1}\\
 &&z^2=z^3=i\label{ziembed2}
\eea
and sections 
\be
\bvX^0= \bvX^1=\hX^0\,,\quad\bvX^2=\bvX^3= \hX^1 \,, \quad
 \bvF_0= \bvF_1=\hF_0\,,\quad\bvF_2=\bvF_3= \hF_1\,.
\ee
The scalar potential of this model is
\be
\hV_g= -\frac{\hg^2}{2} \Bslb 4+\frac{1}{x} +x +\frac{y^2}{x} \Bsrb\,.
\ee
The gauge fields and couplings between the models are related by $ \hg^\Lam=\bvg^\Lam=0$ and
\be\label{gaugeembeddings}
\frac{1}{\sqrt{2}}\hg_0=\bvg_0=\bvg_1\,,\quad \frac{1}{\sqrt{2}}\hg_1=\bvg_2=\bvg_3 \,,\quad \quad \bvA^0=\bvA^1=\frac{1}{\sqrt{2}} \hA^0\,,\quad\quad  \bvA^2=\bvA^3=\frac{1}{\sqrt{2}} \hA^1\,. \non
\ee
For this embedding the dual sections are $\hM_0=-i\hL^1$ and $\hM_1=-i\hL^0$ so that in total we have the following symplectic vector of sections
\be\label{Vtilde}
\bvcV^T=\frac{1}{\sqrt{2}}(\hL^0,\hL^0, \hL^1,\hL^1,-i\hL^1,-i\hL^1,-i\hL^0,-i\hL^0)\,.
\ee

The duality transformation  $\cO_{23} (\al)$ acts trivially  while  $\cO_{12}(\al)$  acts on the sections as follows:
\bea\label{sec12transform}
\bvcV_{\al} &=& \cS\, \cO_{12}(\al) \,\cS^{-1} \bvcV
\eea
where $\cS$ is given in \eq{Sdefinition}. 
From \eq{sec12transform} one can work out that after the transformation we retain the identity $z^2=z^3=i$ but this is also clear since they are fixed points of the fractional linear transformations (\ref{zsol1}-\ref{zsol3}). The scalar field $z^1$ transforms by a fractional linear transformation
\bea\label{z1transform}
z^1_\al &=& \frac{c_\al z^1- s_\al}{s_\al z^1 +c_\al}\,.
\eea
 The new gauge field strengths are obtained from 
\bea
\bpm \bvF^\Lam \\ \bvG_\Lam \epm_{\al} &=& \cS\, \cO_{12}(\al) \, \cS^{-1}\bpm \bvF^\Lam \\ \bvG_\Lam \epm \non
\eea
where we have used the dual field strength defined in \eq{DualFieldStrength} and one finds that this too is invariant. As a result $\cO_{12}$ acts directly on the model of \eq{FX0X1}.

Now we can act on a particular solution such as the black hole of \cite{Klemm:2011xw} in a straightforward manner. This seed solution can be found explicitly in \cite{Klemm:2011xw,Gnecchi:2013mja} which we briefly review and add a few comments regarding the parameter space of this solution. 

The space-time metric for this rotating solution is given by
\bea
ds^2&=&  \frac{\rho^2 -\Delta^2}{\Delta_r}  dr^2 +\frac{\rho^2-\Delta^2}{\Delta_\tha}d\tha^2+ \frac{\Delta_\tha \sinh^2 \tha}{\rho^2-\Delta^2} \blp j dt-(r^2 +j^2-\Delta^2)d\phi \brp^2  \non \\
&&-\frac{\Delta_r}{\rho^2-\Delta^2} \blp dt +j\sinh^2\tha d\phi\brp^2
\eea
and the complex scalar fields are
\bea
&& z^1=-\frac{2j \Delta \cosh\tha}{j^2 \cosh^2\tha +(r-\Delta)^2}+i\,\frac{j^2 \cosh^2\tha+r^2-\Delta^2}{j^2 \cosh^2\tha +(r-\Delta)^2}\,, \\
&& z^2=z^3=i
\eea
where
\bea
\rho^2=r^2 +j^2 \cosh^2\tha\,,\quad\quad \Delta_r=\frac{1}{R^2}\Blp r^2+\frac{j^2-R^2}{2}-\Delta^2 \Brp^2\,,\quad\quad \Delta_\tha= 1+\frac{j^2}{R^2} \cosh^2\tha\,. \non
\eea
The gauge field is given by
\bea
\bvA^\Lam &=& \frac{1}{ 8\bvg}\frac{\cosh\tha}{(\rho^2-\Delta^2)}\blp jdt-(r^2+j^2-\Delta^2)d\phi \brp\,,\quad \Lam=0,1,2,3\,.
\eea
This is a rotating generalization of the solution in section \eq{sec:equalcharges}. The parameter $\Xi$ which appears in \cite{Klemm:2011xw} is unphysical and in our expression has been absorbed by a rescaling of the co-ordinates which appear there. As with the static solution in section \eq{sec:equalcharges} all charges are equal as in \eq{chargesequal}. The parameter $j$ is the rotation parameter, $\Delta$ represents a deformation of the boundary theory by a source.

After setting up these pieces, it is completely straightforward to utilize a non-trivial action of $\cO_{12}(\al)$ on this solution under which the metric, gauge fields and $(z^2,z^3)$ are invariant while $z^1$ transforms exactly as \eq{z1transform}: 
\bea
z^1\ra \frac{c_\al \bslb -2j \Delta \cosh \tha+i(j^2 \cosh^2\tha + r^2 -\Delta^2) \bsrb - s_\al \bslb j^2 \cosh^2 \tha+(r - \Delta)^2 \bsrb}{s_\al \bslb -2j \Delta \cosh \tha+i(j^2 \cosh^2\tha + r^2 -\Delta^2) \bsrb +c_\al \bslb j^2 \cosh^2 \tha+(r - \Delta)^2 \bsrb}\,.
\eea
This results in a family of rotating solutions with rotation parameter $j$ and two additional parameters $(\Delta,\al)$. The discussion below \eq{z1static} is equally valid for this black hole.
When $\Delta=\al=0$ we recover the constant scalar rotating solution of \cite{Caldarelli1999}.

\section{Conclusions}

We have demonstrated that a well-known and simple STU-model of four dimensional gauged supergravity has a powerful and previously un-utilized duality group. The duality group is a property of the theory itself and as such can be used to act on any given solution, we have used this group to generate new classes of supersymmetric AdS$_4$ black holes. 

When acting on the generic supersymmetric static black holes of \cite{Cacciatori:2009iz} we have generated two additional directions in the solution space, both supersymmetric. In the symplectic duality frame in which this directly embeds into the de Wit-Nicolai $\cN=8$ theory, these new directions include two additional electric charges and have non-trivial profiles for the axions. One particular representative of our new solutions had been previously constructed numerically in \cite{Halmagyi:2013qoa}. Using the results of \cite{Halmagyi:2013qoa} for the static BPS horizon geometries in $\cN=2$ $U(1)$-gauged supergravity theories, we have conjectured that with the new results of this paper in hand, the known solution space of supersymmetric static black holes in the STU-model is now co-dimension one within the full space of solutions. The sixth and final dimension of the solution space remains undiscovered and we predict that it should involve a non-trivial profile for the phase of the supersymmetry parameter, much like the quite complicated supersymmetric static black holes with hypermulitplets found in \cite{Halmagyi:2013sla}. We have not presented a strategy by which one could use duality to generate this final branch but one could surely use numerics to confirm its existence.

When acting on the black holes of \cite{Cacciatori:2009iz} with equal charges, we have generated a new parameter in the solution which corresponds to additional scalar hair. This black hole now has two free parameters, one is dual to an explicit mass term in the world-volume M2-brane theory. This is inaddition to the mass terms induced from twisting of the theory and the curvature couplings \cite{Maldacena:2000mw}. The new parameter we have generated must then correspond to a vev.

We have also used the duality group to generate supersymmetric rotating black holes by using the rotating black hole of Klemm \cite{Klemm:2011xw} as a seed solution. This family remains within the $\hF=-i\hX^0\hX^1$ model but to generate this family we had to first embed this model into the STU-model.  Our new solutions have one additional parameter with respect to the Klemm black hole corresponding to additional scalar hair. In the recent work \cite{Chow:2013gba} a new family of rotating AdS$_4$ black holes was found by explicitly solving the second order field equations, generalizing the work of \cite{Kostelecky:1995ei, Cvetic:2005zi}. The Killing spinor conditions were not checked in that work and they do not reference \cite{Klemm:2011xw} but it would certainly be interesting to establish whether there is overlap between our results in section \ref{sec:GenerateRotation} and the results of \cite{Chow:2013gba}. The supersymmetric black hole of \cite{Kostelecky:1995ei} and its generalizations have a lower bound on the angular momentum whereas the rotating black holes of section \ref{sec:GenerateRotation} have a regular static limit. There is clearly more work to be done regarding supersymmetric AdS$_4$ black holes even in the STU model; there remains the open problem of constructing a supersymmetric rotating black hole which has a regular CK black hole with $S^2$ horizon as its zero-rotation limit.

There has been much recent work developing non-BPS black holes in gauged supergravity \cite{Gnecchi:2013mja, Chow:2013gba,Klemm:2012yg, Gnecchi:2012kb, Hristov:2013sya} and one can straightforwardly use our duality group on these as well. For non-BPS black holes which are finite temperature generalizations of the CK black holes, one would expect to find qualitatively similar results to ours. The space of static non-BPS solutions discussed in \cite{Chow:2013gba} has no overlap with our solution space of supersymmetric black holes in section \ref{sec:GenerateStatic} but it would appear that our duality transformations would not generate new solutions in the class of static black holes found in \cite{Chow:2013gba} since in that class all charges are already accounted for. Nonetheless it would be interesting to check this in detail.

Our solution generating technique is reminiscent of the TST duality \cite{Lunin:2005jy} used in the study of AdS solutions of IIB and eleven-dimensional supergravity. In that work, families of AdS solutions were generated which correspond to the gravity dual of the deformation of the superconformal field theory by exactly marginal operators. This is clearly not directly related to our duality group since the de Wit-Nicolai theory (of which our STU-model is a truncation) contains the AdS$_4$ scalars dual to {\it relevant} operators, nonetheless we find it an interesting point of comparison.  While Lunin-Maldacena focused on BPS solutions, using the techniques of \cite{Lunin:2005jy} one can find additional non-BPS directions in the solution space \cite{Frolov:2005dj}. Like the generator $\cO_{g}(\tha)$  in section \ref{sec:thirdgen}, these resulted from dualizing along directions where the bosonic fields are neutral but the Killing spinor is charged. For solutions of IIB supergravity which are topologically of the form AdS$_5\times S^5$, the solution space is conjectured to admit an additional direction%
\footnote{This search for the resulting supergravity solution remains a long-standing open problem, the state of the art in perturbation theory can be found in \cite{Aharony:2002hx}.} 
\cite{Leigh:1995ep} than that found in \cite{Lunin:2005jy}. This is the dual of the so-called {\it cubic} deformation of $\cN=4$ SYM and cannot be obtained in any known way through duality. If finding the exact supergravity solution for the final direction of our conjectured solution space of static BPS black holes is a problem of comparable difficulty, one should note that this would be quite a formidable problem.

Duality in gauged supergravity has rarely been employed in the literature. An attempt to use the Geroch group in reductions to three dimensions was carried out in \cite{Berkooz2008} but such a method has not yet proved as useful for generating rotating black holes as it is for ungauged supergravity. It is possible that our results for these $\cN=2$ U(1)-gauged supergravity theories could help in this regard, certainly it should be possibly to understand duality for black holes with hyeprmultiplets \cite{Halmagyi:2013sla}. More generally we hope and expect that the synthesis of our new duality techniques with the numerous recents works on black holes in gauged supergravity will result in much further progress in the study of asymptotically AdS black holes.

\vskip 10mm
\noindent {\bf Acknowledgements:} We would like to thank Guillaume Bossard and Chiara Toldo for interesting discussions, in particular Alessandra Gnecchi and Michela Petrini for collaborations at an early stage of this project.
\vskip 10mm

\begin{appendix}

\section{Special geometry}\label{app:SpGeometry}

This material is all standard but we include it to make our conventions clear and in particular to be straight with our numerical factors. We essentially use the $\cN=2$ supergravity conventions of \cite{Andrianopoli:1996vr} except we use the mostly plus signature $(-+++)$. The supergravity action is given by
\bea\label{sugraaction}
S_{4d}=\int\,d^4 x\sqrt{-g} \Blp \frac{1}{2}R- g_{ij} \del_{\mu} z^i\del^{\mu} z^j + \cI_{\Lambda\Sigma}F^{\Lambda}_{\mu \nu}F^{\Sigma\,\mu \nu}+ \cR_{\Lam \Sig} F_{\mu\nu}^\Lam( \half\eps^{\mu\nu\rho\sig} F_{\rho\sig}^\Sig)-V_g\Brp
\eea
The prepotential we use is
\bea
F&=& -d_{ijk} \frac{X^i X^j X^k}{X^0} \label{Prepotential}
\eea
and special co-ordinates are
\bea
X^\Lam&=& \bpm1 \\ z^i\epm\,,\quad\quad z^i = x^i + i y^i\,.
\eea
From this we obtain that the dual sections $F_\Lam=\del_\Lam F$ are
\bea
F_\Lam&=& \bpm d_{ijk} z^i z^j z^k \\ -3 d_{z,i} \epm \label{FLamexplicit}
\eea
and the K\"ahler potential and metric are
\bea
e^{-K}&=& 8  d_y\,,\quad\quad g_{i\jbar} =\del_i \del_{\jbar} K\,.
\eea
As usual the rescaled sections are defined as
\be\label{LMdef}
\cV=\bpm L^\Lam \\ M_\Lam \epm=e^{K/2}\bpm  X^\Lam \\ F_\Lam\epm \,.
\ee
The kinetic and topological terms for the vector fields in \eq{sugraaction} come from the tensor
\bea
\cN_{\Lam\Sig}&=&\cR_{\Lam\Sig}+i\, \cI_{\Lam\Sig }= \Fbar_{\Lam\Sig} + 2i \frac{\Im F_{\Lam \Delta} \Im F_{\Sig \Upsilon} X^\Delta X^\Upsilon }{ \Im F_{\Delta \Upsilon} X^\Delta X^\Upsilon}
\eea
where $F_{\Lam\Sig}=\del_\Lam \del_\Sig F$. The dual gauge-field strength is 
\be\label{DualFieldStrength}
G_\Lam= \cR_{\Lam \Sig} F^\Sig - \cI_{\Lam\Sig} * F^\Sig\,.
\ee

We will also use the following tensor%
\footnote{The hat index here does not refer to any particular duality frame, hopefully this does not cause confusion on the part of the reader.}
\bea
\hd^{ijk} =\frac{g^{il}g^{jm} g^{kn} d_{ijk}}{d_y^2}
\eea
which has the crucial property that it is constant whenever $\cM_v$ is a homogeneous space.
We use the following shorthand for contraction of objects with the symmetric tensors $d_{ijk}$ and $\hd^{ijk}$:
\bea
&&d_g=d_{ijk} g^i g^j g^k\,,\ \ \ d_{g,i}= d_{ijk} g^j g^k\,,\ \ \ \ d_{g,ij}=d_{ijk} g^k\,, \non \\
&&\hd_g=\hd^{ijk} g_i g_j g_k\,,\ \ \ \hd_{g}^{i}= \hd^{ijk} g_j g_k\,,\ \ \ \ \hd_g^{ij}=\hd^{ijk} g_k\,.
\eea

At various points in the text we have used different symplectic frames. For example we have four different sections $L^\Lam$
\bea
L^\Lam:&& {\rm sections\ in\ the\ STU\ model\ with\ cubic\ prepotential,\ see\ eq.}\ \eq{FSTU}\non \\
\bvL^{\Lam}:&& {\rm sections\ of\ STU\ model\ in\ frame\ with}\ \bvF=-2i\sqrt{\bvX^0\bvX^1\bvX^2\bvX^3}\,,\ {\rm see\ eq.}\ \non \eq{SympFrame1} \\
\tL^\Lam:&& {\rm sections}\ L^\Lam\ {\rm rescaled\ by\ a\ metric\ factor,\ see\ eq.}\ \eq{rescaledLM} \non\\
\hL^\Lam:&& {\rm sections\ in\ the\ model\ with}\ \hF=-i\hX^0\hX^1\,,\ {\rm see\ eq.}\ \eq{FX0X1} \non
\eea

\section{Duality symmetries and Very Special K\"ahler Geometry}\label{app:Duality}
We now summarize some key aspects of duality symmetries for very special K\"ahler geometry following \cite{deWit:1991nm,deWit:1992wf, deWit:1993rr}. Under the action of $Sp(2n_v+2,\RR)$, the prepotential transforms according to 
\bea
\cS&=& \bpm A& B \\ C & D\epm \in Sp(2n_v+2,\RR)\,, \\
\tF(\tX)&=&F(X)+X^\Lam (C^t B)_\Lam^{\ \Sig} F_{\Sig}+\frac{1}{2} X^\Lam (C^t A)_{\Lam\Sig}  X^{\Sig} +\frac{1}{2} F_\Lam (D^t B)^{\Lam\Sig } F_\Sig\,.
\eea

The elements of $Sp(2n_v+2,\RR)$ which leave the prepotential invariant correspond to isometries of $\cM_v$ and these have been classified by de Wit and Van-Proeyen. Working at the level of the Lie algebra we have an element
\bea\label{Sunderline} 
\underline{\cS}&=& \bpm Q& R\\ S & T \epm \in \sp(2n_v+2,\RR)
\eea
with components
\bea
Q&=&-T^t = \bpm \beta &  a_i  \\ b^j &  B^{i}_{\ j}+\frac{1}{3}  \beta \delta^i_j \epm\,, \label{QTmatrices}\\
R&=& \bpm 0 & 0 \\ 0 & -\frac{3}{32}\hd^{ijk} a_k \epm\,, \label{Rmatrix}\\
S&=& \bpm 0 & 0 \\ 0 & -6 d_{ijk} b^k \epm  \label{Smatrix}\,.
\eea
The scalar fields transform infinitesimally as
\bea
\delta z^i &=& b^i -\frac{2}{3} \beta z^i + B^{i}_{\  j}z^j - \frac{1}{2} R^{i\ \ l}_{\ jk} z^j z^k a_l\,.
\eea
where $R^{i\ \ l}_{\ jk} $ is the Riemann tensor on $\cM_v$:
\be
R^{i\ \ l}_{\ jk} =2\delta^i_{(j}\delta^l_{k)} -\frac{9}{16} \hd^{ilm} d_{mjk}\,.
\ee
In general these symmetries are constrained
\bea
B^{i}_{\ (j} d_{kl) i}&=& 0\,, \\
a_{i} E^{i}_{jklm} &=& 0 \label{aConstraint}
\eea
where the $E$-tensor is given by
\bea
E^{i}_{jklm}&=&  \hd^{inp}d_{n(jk} d_{lm)p}-\frac{64}{27} \delta^i_{(j}d_{klm)}\,.
\eea
When $\cM_v$ is a homogeneous space, the case of most interest to us, $E^i_{jklm}$ vanishes and thus the constraint \eq{aConstraint} is identically zero. As a consqequence the $a_i$ and $b^j$ parameters are unconstrained.

To get a feeling for these symmetries, consider the fractional linear transformation of $z^i$ under $SL(2,\RR)$. To work out the infinitesimal tranformation we take the standard generators of $\sl(2,\RR)$
\be
E=\bpm 0& 1 \\ 0 & 0\epm\,,\ \ \ 
F=\bpm 0& 0 \\ 1 & 0\epm\,,\  \ \ \
H=\bpm 1& 0 \\ 0 & -1 \epm
\ee
then we have
\bea
\delta_E z^i\ra \al \,,\ \ \ \ \ \delta_F z^i\ra -\al (z^i)^2\,,\ \ \ \ \ 
\delta_H z^i \ra  2\al z^i\,.
\eea
So one can interpret the matrix $\underline{\cS}$ in \eq{QTmatrices}-\eq{Smatrix} with $b^i\neq 0$ as raising operators and when $\cM_v$ is a homogeneous space, the Riemann tensor is constant and one can interpret the matrix with $a_i\neq0$ as lowering operators. The $(\beta, B^{i}_{\ j})$ are then the Cartan elements. The full commutation relations can be easily worked out or found in \cite{deWit:1991nm,deWit:1992wf, deWit:1993rr}.

\end{appendix}

\providecommand{\href}[2]{#2}\begingroup\raggedright\endgroup

\end{document}